\documentclass[11pt,reqno]{amsart}

\usepackage{graphicx}
\usepackage{amsmath,amssymb}

\usepackage{xcolor}

\newcommand{\be}{\begin{equation}}
\newcommand{\ee}{\end{equation}}

\newcommand{\mk}{\langle k \rangle}
\newcommand{\bk}{{\bf k}}
\newcommand{\erf}{\mbox{erf}}

\begin{document}

\title{Explosive behaviour in networks of Winfree oscillators}

\author{Shawn Means}
\address{School of Natural and Mathematical Sciences, 
Massey University,
Private Bag 102-904 
North Shore Mail Centre, 
Auckland 0745,
New Zealand
}%
\email{s.means@massey.ac.nz}

\author{Carlo R. Laing}
\address{School of Natural and Mathematical Sciences, 
Massey University,
Private Bag 102-904 
North Shore Mail Centre, 
Auckland 0745,
New Zealand
}%
 \email{c.r.laing@massey.ac.nz}

\begin{abstract}
We consider directed networks of Winfree oscillators with power law distributed in- and
out-degree distributions. Gaussian and power law distributed intrinsic frequencies are considered,
and these frequencies are positively correlated with oscillators' in-degrees. The Ott/Antonsen
ansatz is used to derive degree-based mean field equations for the expected dynamics of networks,
and these are numerically analysed. In a variety of cases ``explosive''
transitions between either two different steady states or between a steady state and a periodic
solution are found, and these transitions are explained using bifurcation theory.

\end{abstract}

\maketitle

\section{Introduction}
The study of networks is driven by the many real-world systems which can be represented
in this way~\cite{newman2003}, such as networks of neuronal~\cite{pernice2011}, cardiac~\cite{casaleggio2014}
or smooth muscle~\cite{xu2015} cells. In networks of oscillators it is often the conditions
under which the oscillators synchronise which are 
of interest~\cite{pikros01,str00}. For networks of nonidentical
oscillators there is often a continuous transition from asynchrony to higher and higher
levels of synchrony as the strength of coupling between oscillators is 
increased~\cite{kuramoto1984,str00}. But recently a different form of transition known
as ``explosive synchrony'' (ES) has been 
observed. Here, there is
a discontinuous increase in network coherence as the coupling strength between oscillators is increased (see~\cite{dsouza2019} for a review). This behaviour was first reported by~\cite{gomez2011} 
in an undirected network of Kuramoto oscillators~\cite{kuramoto1984} for which the intrinsic
frequency of an oscillator was equal to its degree, the degrees having an inverse power law
distribution. The authors emphasised that the explosive transition was due to there being
a relationship
between a local network property (an oscillator's degree) and a property of an oscillator
(its intrinsic frequency) --- in this case a positive correlation between the two.
Since the publication of~\cite{gomez2011} there have been many studies of 
ES~\cite{leynav13,bocalm16,zhazou14,zhaboc15}.

As recently noted by~\cite{kuebic21}, ES is not actually an unexpected phenomenon, since if there
is a continuous transition to synchrony through either a transcritical or pitchfork bifurcation
as, say, the strength of coupling between oscillators is increased, then by varying a second parameter
sufficiently the criticality of the bifurcation will change. This change leads to bistability and
a region of hysteresis as the original parameter is varied (i.e.~an ``explosive'' transition).

Many studies of ES use Kuramoto phase oscillators which are coupled through phase differences.
However, more realistic oscillators are not coupled in this way, instead having interactions
which depend on the states of the two oscillators which are 
coupled~\cite{tsuter07,moocoo15,borkop03,lukbar13}.
The oscillator we study -- the Winfree model~\cite{winfree67} -- 
is a phase oscillator similar in form to the Kuramoto 
oscillator 
but is not described with phase differences, instead involving a ``phase response curve'' and a pulsatile
function of phase. We are not aware of any reports of ES in networks of Winfree oscillators, and
as one of our contributions we report 
here our observations of this phenomenon in our numerical explorations of
networks of Winfree oscillators in a variety of network configurations. 

Most previous studies of networks of Winfree oscillators have focused on the all-to-all coupled 
case~\cite{ariaratnam2001,pazmon14,ha2015,gallego2017,pazmon19}, with an exception being \cite{laing2021}. Many previous studies of ES have considered undirected networks,
where oscillators are either coupled or not. However, we consider directed networks -- as seen in neuronal systems -- where one oscillator influences other oscillators coupled downstream but without necessarily a reciprocal influence back upstream. 

The bistability that often accompanies ES is of interest as it allows a network to have
more than one stable state for a given set of parameters, with the current state often
determined by the network's history. A prominent example of such bistability is
in networks which exhibit Up and Down states~\cite{wilkaw96,plekit98}, with transitions between
these states driven by either fluctuations or noise of some form, or by transient inputs.
The abrupt transitions between states may also provide a model for
the conscious-unconscious transition when awaking from
anesthesia~\cite{steste99}.

The structure of the paper is as follows. In Sec.~\ref{sec:model} we present the network
we consider and the degree-based mean field description of its dynamics, derived using
the Ott/Antonsen ansatz. Specific cases of independent in- and out-degrees, and equal
in- and out-degrees, are considered. In Sec.~\ref{sec:gauss} we consider a Gaussian
distribution of intrinsic frequencies which are positively correlated with oscillators'
in-degrees. In Sec.~\ref{sec:pow} we consider an inverse power law distribution of
intrinsic frequencies and again correlate them with oscillators'
in-degrees. We consider the cases of independent in- and out-degrees, and equal
in- and out-degrees. A variety of ``explosive'' transitions are observed and
explained using simple bifurcation theory. We conclude in Sec.~\ref{sec:disc}.


\section{Model}
\label{sec:model}
The model describing a directed network of Winfree oscillators is as 
presented in~\cite{laing2021,pazmon14}:
\be
   \frac{d\theta_j}{dt}=\omega_j+U(\theta_j)\frac{\epsilon}{\mk}\sum_{n=1}^NA_{jn}T(\theta_n); \qquad j=1,2\dots N  \label{eq:dthdt}
\ee
where $\epsilon$ is the strength of coupling, $\mk$ is the mean degree of the network and
the connectivity is given by the adjacency matrix $A$ with $A_{jn}=1$ if oscillator $n$ connects
to oscillator $j$ and zero otherwise. The $\omega_j$ are chosen from a Lorentzian
distribution with centre and half-width at half-maximum to be discussed below.
The phase response curve~\cite{schpri11,netban05} is chosen to be $U(\theta)=-\sin{\theta}$ and the 
pulsatile function $T$ which has a maximum at $\theta=0$ is
\be
   T(\theta)=\frac{8}{35}(1+\cos{\theta})^4
\ee
The in-degree of oscillator $j$ is
\be
   k_{in,j}=\sum_{n=1}^N A_{jn}
\ee
and the out-degree of oscillator $n$ is
\be
   k_{out,n}=\sum_{j=1}^N A_{jn}
\ee

We consider large networks where all in- and out-degrees are large and assume that the network
can be characterised by its degree distribution $P(\bk)$ where $\bk=(k_{in},k_{out})$,
and an assortativity function $a(\bk'\to \bk)$ giving the probability that an oscillator with 
degree $\bk'$ connects to one with degree $\bk$, given that such oscillators exist.
$P(\bk)$ is normalised such that $\sum_{\bk}P(\bk)=N$ and we choose the marginal distributions
of the in-degrees and the out-degrees to be equal.

Using the theory in~\cite{laing2021} (or see~\cite{laibla20,chahat17} for similar derivations)
one can show that the long-time dynamics of the network
is described by
\be
   \frac{\partial b(\bk,t)}{\partial t}=\frac{\epsilon R(\bk,t)}{2}+[i\omega_0(\bk)-\Delta(\bk)]b(\bk,t)-\frac{\epsilon R(\bk,t)}{2}[b(\bk,t)]^2
\ee
where $\omega_0(\bk)$ and $\Delta(\bk)$ are the centre and half-width at half-maximum, respectively,
of the Lorentzian distribution from which the values of $\omega_j$ for oscillators with degree
$\bk$ are chosen:
\be
   g(\omega(\bk))=\frac{\Delta(\bk)/\pi}{[\omega(\bk)-\omega_0(\bk)]^2+\Delta^2(\bk)} \label{eq:lor}
\ee

The variable
\be
   b(\bk,t)=\int_{-\infty}^\infty\int_0^{2\pi} f(\theta,\omega|\bk,t)e^{-i\theta}d\theta\ d\omega
\ee
is the complex-valued order parameter for oscillators with degree $\bk$, where 
$f(\theta,\omega|\bk,t)d\theta d\omega$ is the probability that an oscillator with degree $\bk$
has phase in $[\theta,\theta+d\theta]$ and frequency in $[\omega,\omega+d\omega]$ at time $t$.
$R(\bk,t)$ is given by
\be
   R(\bk,t)=\frac{1}{\mk}\sum_{\bk'}P(\bk')a(\bk'\to\bk)G(\bk',t)
\ee
where
\begin{eqnarray}
   G(\bk,t) & = & 1+\frac{4\left[b(\bk,t)+\overline{b}(\bk,t)\right]}{5}+\frac{2\left[b^2(\bk,t)+\overline{b}^2(\bk,t)\right]}{5} \nonumber \\
& & +\frac{4\left[b^3(\bk,t)+\overline{b}^3(\bk,t)\right]}{35}+\frac{b^4(\bk,t)+\overline{b}^4(\bk,t)}{70} \label{eq:G}
\end{eqnarray}
where overline indicates the complex conjugate. The form of $G$ is determined by the
function $T(\theta)$. This derivation of a degree-dependent
mean field description uses the Ott/Antonsen ansatz~\cite{ottant08,ottant09}. A crucial
ingredient for the use of this ansatz is the sinusoidal form of the function $U(\theta)$.

We are interested in the case of neutral assortativity, for which~\cite{resott14}
\be
   a(\bk'\to\bk)=\frac{k_{out}'k_{in}}{N\mk}
\ee
and cases where either the in- and out-degree of an oscillator are independent, or they are equal
(and thus perfectly correlated).
Writing $P(k_{in}',k_{out}')$ instead of $P(\bk')/N$ we have
\be
   R(k_{in},k_{out})=\frac{k_{in}}{\mk^2}\sum_{k_{in}'}\sum_{k_{out}'}P(k_{in}',k_{out}')k_{out}'G(k_{in}',k_{out}',t)
\ee
which is clearly independent of $k_{out}$. Thus both $b$ and $G$ must also be independent of $k_{out}$
and we have
\be
  R(k_{in},t)=\frac{k_{in}}{\mk^2}\sum_{k_{in}'}Q(k_{in}')G(k_{in}',t)
\ee
where
\be
   Q(k_{in}')=\sum_{k_{out}'}P(k_{in}',k_{out}')k_{out}'.
\ee

For the case of the in- and out-degrees of an oscillator being independent, $P(k_{in}',k_{out}')$
factorises: $P(k_{in}',k_{out}')=p(k_{in})p(k_{out}')$ where $p$ is the marginal distribution 
of either the in- or out-degrees. In this case,
\be
  R(k_{in},t)=\frac{k_{in}}{\mk}\sum_{k_{in}'}p(k_{in}')G(k_{in}',t). \label{eq:Rind}
\ee
Alternatively, if $k_{in}=k_{out}$ for each oscillator (i.e.~each oscillator has the
same in- and out-degree) then $Q(k_{in}')=k_{in}'p(k_{in}')$ and
\be
  R(k_{in},t)=\frac{k_{in}}{\mk^2}\sum_{k_{in}'}p(k_{in}')k_{in}' G(k_{in}',t). \label{eq:Rsame}
\ee
In either case the ``input'' to an oscillator with in-degree $k_{in}$ is proportional to 
its in-degree, i.e.~$k_{in}$.

One could use this theory to investigate the effects of choosing different degree distributions
$p$ as in~\cite{lai21} but instead we fix $p$ as a power law distribution and look at the
effects of correlating an oscillator's frequency with its degree. Specifically, we make 
$\Delta$ independent of degree but choose $\omega_0$ to be a function of $k_{in}$ only.
(One could make $\omega_0$ a function of $k_{out}$ as well, or instead, but doing so increases
the computational cost, and the phenomena we are interested in occur when $\omega_0$ is 
a function of $k_{in}$ only.) Thus the equations of interest are
\be
   \frac{\partial b(k_{in},t)}{\partial t}=\frac{\epsilon R(k_{in},t)}{2}+[i\omega_0(k_{in})-\Delta]b(k_{in},t)-\frac{\epsilon R(k_{in},t)}{2}[b(k_{in},t)]^2 \label{eq:dbdt}
\ee
where $m\leq k_{in}\leq M$, where $m$ and $M$ are the minimum and maximum in- and out-degrees
respectively, $R$ is given by either~\eqref{eq:Rind} or~\eqref{eq:Rsame} and $G$ is given
by~\eqref{eq:G} but with only dependence on in-degree.

Note that in previous work~\cite{laing2021} we used similar equations as those above
to investigate the effects of varying the
correlation between in- and out-degrees of Winfree oscillators, and of parameter- and
degree-based assortativity.
We also examined the effects of correlating 
oscillators' intrinsic frequencies with their in-degrees, but did not consider varying
the coupling strength. Also, there we considered uniform degree distributions rather than power law,
as we do here and as many other researchers do.

Below we will use the mean firing frequency of a network to describe its behaviour, so now 
derive an expression for that. The dynamics of an oscillator with in-degree $k_{in}$ is
\be
   \frac{d\theta}{dt}=\omega(k_{in})-\epsilon R(k_{in})\sin{\theta}.
\ee
If $|\omega(k_{in})|>\epsilon R(k_{in})$ then the oscillator will fire periodically with
frequency 
\be
  \frac{\sqrt{\omega^2(k_{in})-\epsilon^2 R^2(k_{in})}}{2\pi}.
\ee
Thus the expected frequency for oscillators with in-degree $k_{in}$ is
\begin{align}
   f(k_{in}) & = \frac{1}{2\pi}\mbox{Re}\left(\int_{-\infty}^\infty g(\omega(k_{in}))\sqrt{\omega^2(k_{in})-\epsilon^2 R^2(k_{in})}\ d\omega(k_{in})\right) \nonumber \\
      & = \frac{1}{2\pi}\sqrt{\frac{\omega_0^2(k_{in})-\Delta^2-\epsilon^2 R^2(k_{in})+\sqrt{[\omega_0^2(k_{in})-\Delta^2-\epsilon^2 R^2(k_{in})]^2+4\Delta^2\omega_0^2(k_{in})}}{2}}
\label{eq:fkin}
\end{align}
where $g$ is the Lorentzian~\eqref{eq:lor} 
and thus the mean firing rate over the network is 
\be
   F=\frac{1}{2\pi}\sum_{k_{in}}p(k_{in})f(k_{in})
\ee

For comparison with our results below we briefly describe the dynamics of a fully connected
network of Winfree oscillators. Two types of behaviour are typically seen in such a network: synchronous and asynchronous~\cite{pazmon14,gallego2017}, 
although the fraction of oscillators actually oscillating varies in different asynchronous states. Increasing $\epsilon$ tends to destroy synchronous behaviour through a saddle-node-on-invariant-circle (SNIC) bifurcation, as many of the oscillators ``lock'' to 
an approximate fixed point. For moderate $\epsilon$ increasing the spread of intrinsic frequencies 
tends to destroy synchronous behaviour through a supercritical Hopf bifurcation, as the oscillators become too dissimilar in frequency to synchronise~\cite{pazmon14}. Below we will see a wider
variety of bifurcations resulting from the networks' structure.

\section{Gaussian frequency distribution}
\label{sec:gauss}
We choose the degree distribution $p(k)$ to be a truncated power law distribution with exponent $-3$,
as many others have done when studying ES~\cite{liu2013,gomez2011}:
\be
   p(k)=\begin{cases} a/k^3 & m\leq k\leq M \\ 0 & \mbox{ otherwise} \end{cases} \label{eq:pk}
\ee 
where $a$ is a normalisation such that
\be
   \sum_{k=m}^M \frac{a}{k^3} =1. \label{eq:norm}
\ee
Since the degrees are all large (i.e.~$1\ll m$) we treat $k$ as a continuous variable and 
approximate the sum in~\eqref{eq:norm} by an integral, giving $a=2m^2M^2/(M^2-m^2)$. The 
cumulative distribution function for $k$ is
\be
   \widehat{p}(k)=\int_m^k p(s)\ ds=\frac{a}{2}\left(\frac{1}{m^2}-\frac{1}{k^2}\right).
\ee
We need to specify the
dependence of $\omega_0$ on $k_{in}$. In this section we consider the case where for a particular
realisation of the network, we randomly choose 
the {\em target} frequencies from a Gaussian distribution with
mean 1 and standard deviation $\sigma$, then assign the smallest target frequency to the oscillator
with smallest in-degree, the second smallest target frequency to the oscillator with second smallest
in-degree, etc. (For oscillators with equal in-degree, they are ranked in random order.)
For oscillator $j$ the {\em actual} $\omega_j$ is then chosen from a 
Lorentzian with centre equal to the target frequency for oscillator
$j$ and half-width at half-maximum (HWHM) equal to $\Delta$, similar to the scheme in~\cite{skares15}. 

In this case we have
\be
   \omega_0(k)=\widehat{q}^{-1}(\widehat{p}(k)) \label{eq:om0k}
\ee
where $\widehat{q}$ is the cumulative distribution function of the appropriate Gaussian distribution,
i.e.
\be
   \widehat{q}(\omega)=\frac{1}{2}\left[1+\erf\left(\frac{\omega-1}{\sqrt{2}\sigma}\right)\right]
\ee

\begin{figure}
\begin{center}
\includegraphics[width=14cm]{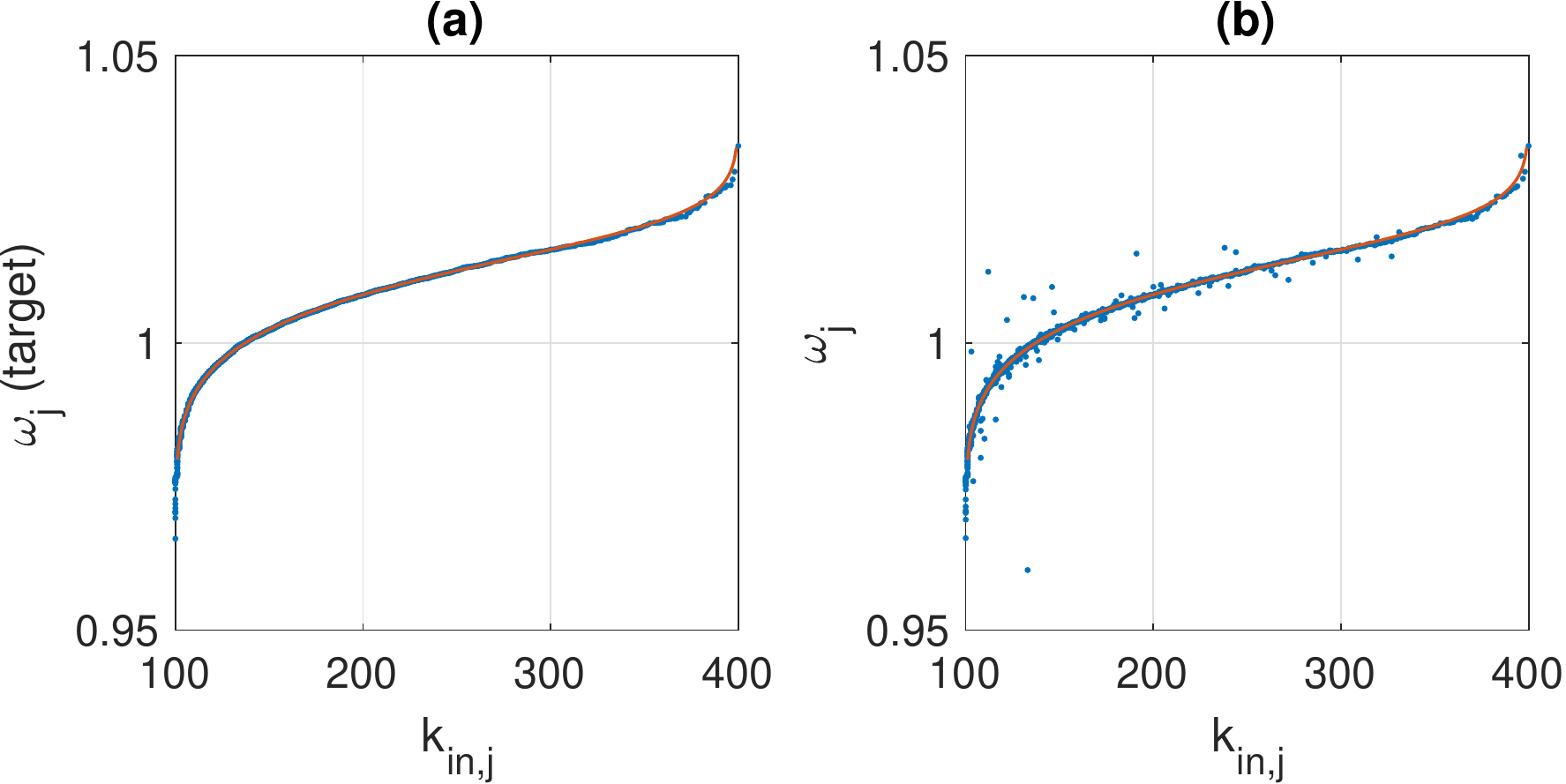}
\caption{(a): solid line:~\eqref{eq:om0k}; dots: target frequencies and degrees for one realisation
of a network. (b): solid line:~\eqref{eq:om0k}; dots: actual frequencies and degrees for 
one realisation of a network. Parameters: $m=100,M=400, N=2000, \sigma=0.01,\Delta=0.00005$.}
\label{fig:dist}
\end{center}
\end{figure}

A demonstration of this is shown in Fig.~\ref{fig:dist} where we create a directed network
with $N=2000$ nodes using the configuration model~\cite{newman2003}, then randomly chose $N$
target frequencies from a Gaussian distribution with mean 1 and standard deviation $0.01$,
then associated the smallest frequency with the node with smallest in-degree etc. These values
are shown as dots in panel (a), along with the theoretical relationship~\eqref{eq:om0k}. Panel (b)
shows the actual values of the $\omega_j$ (dots), taken from a Lorentzian with centre equal to the
values in panel (a) and HWHM 0.00005.

Clearly there is a positive correlation between $k_{in,j}$ and $\omega_j$, but 
the Pearson correlation coefficient between them, defined by
\be
    \rho_{k\omega}=\frac{\sum_{j=1}^N (k_{in,j}-\mk)(\omega_j-\bar{\omega})}{\sqrt{\sum_{j=1}^N(k_{in,j}-\mk)^2}\sqrt{\sum_{j=1}^N(\omega_j-\bar{\omega})^2}}
\ee
where $\bar{\omega}$ is the mean of the $\omega_j$,
is less than 1 and (using this construction)
cannot be systematically
varied, as was possible in~\cite{laing2021,laibla20}. Note that the idea of not having a
perfect match between an oscillator's degree and a prescribed frequency (as we have here
for $\Delta\neq 0$) was discussed in~\cite{skaare14}, where it was shown that such a mismatch
actually {\em created} ES.

To investigate the influence of varying the correlation between $k_{in,j}$ and $\omega_j$, we 
created a sequence of the appropriate degrees, chose values of $\omega$ from a Gaussian
distribution and randomly paired them. We then repeatedly performed Monte Carlo swaps of $\omega$ values; potential swaps were accepted if they increased the Pearson correlation coefficient toward a target value -- typically a positive number. This approach required substantial effort to satisfy high correlations $(\geq 0.9)$, since a sequence of $\omega$ randomly paired with degree sequences typically exhibited no correlation. Alternatively, we maximised the correlation between the
$k_{in,j}$ and $\omega_j$ by initially sorting both sequences as above. Aligning maximal-minimal values then produced  the highest possible correlation coefficient for given sequences which we then reduced using Monte Carlo swaps, with swaps accepted if they pushed the correlation value toward a target value. Sequences of degrees and frequencies generated with this approach were assembled into adjacency matrices utilising our network assembly scheme, the ``permutation method'', presented previously~\cite{means2020}. Note that with this approach of constructing networks 
the Ott/Antonsen approach cannot
be used, and we must simulate the resulting networks to determine their behaviour.

\subsection{Results}

We numerically investigate~\eqref{eq:dbdt} with~\eqref{eq:Rind}. We evaluate functions
at all integer in-degrees
satisfying $m+1\leq k_{in}\leq M-1$ to avoid the singularities in $\widehat{q}^{-1}$ when its
argument is either 0 or 1, resulting in a moderately large set of ordinary differential equations.
We could use a more efficient method which approximates the sum in~\eqref{eq:Rind} with fewer
``virtual'' degrees as explained in~\cite{laibla20}, but that is unnecessary here. 
Typically we integrate~\eqref{eq:dbdt} to a stable fixed point and then use pseudo-arclength
continuation to follow the fixed point as parameters are varied, determining the
stability of the fixed point from the eigenvalues of the linearisation of the dynamics
about the fixed point~\cite{lai14B,gov00}. Periodic solutions are studied in a similar
way by putting a Poincar{\'e} section in the flow (at Re[$b(m+1,t)]=0$) and integrating from this
section until the solution next hits this section. Stability is given by the Floquet
multipliers of the periodic solution.

The usual complex-valued order parameter defined for the network~\eqref{eq:dthdt} is
\be
   Y(t)=\frac{1}{N}\sum_{j=1}^N e^{i\theta_j}
\ee
and for~\eqref{eq:dbdt} the appropriate measure is
\be
   Z(t)=\sum_{k_{in}}p(k_{in})\bar{b}(k_{in},t)
\ee

\begin{figure}
\begin{center}
\includegraphics[width=14cm]{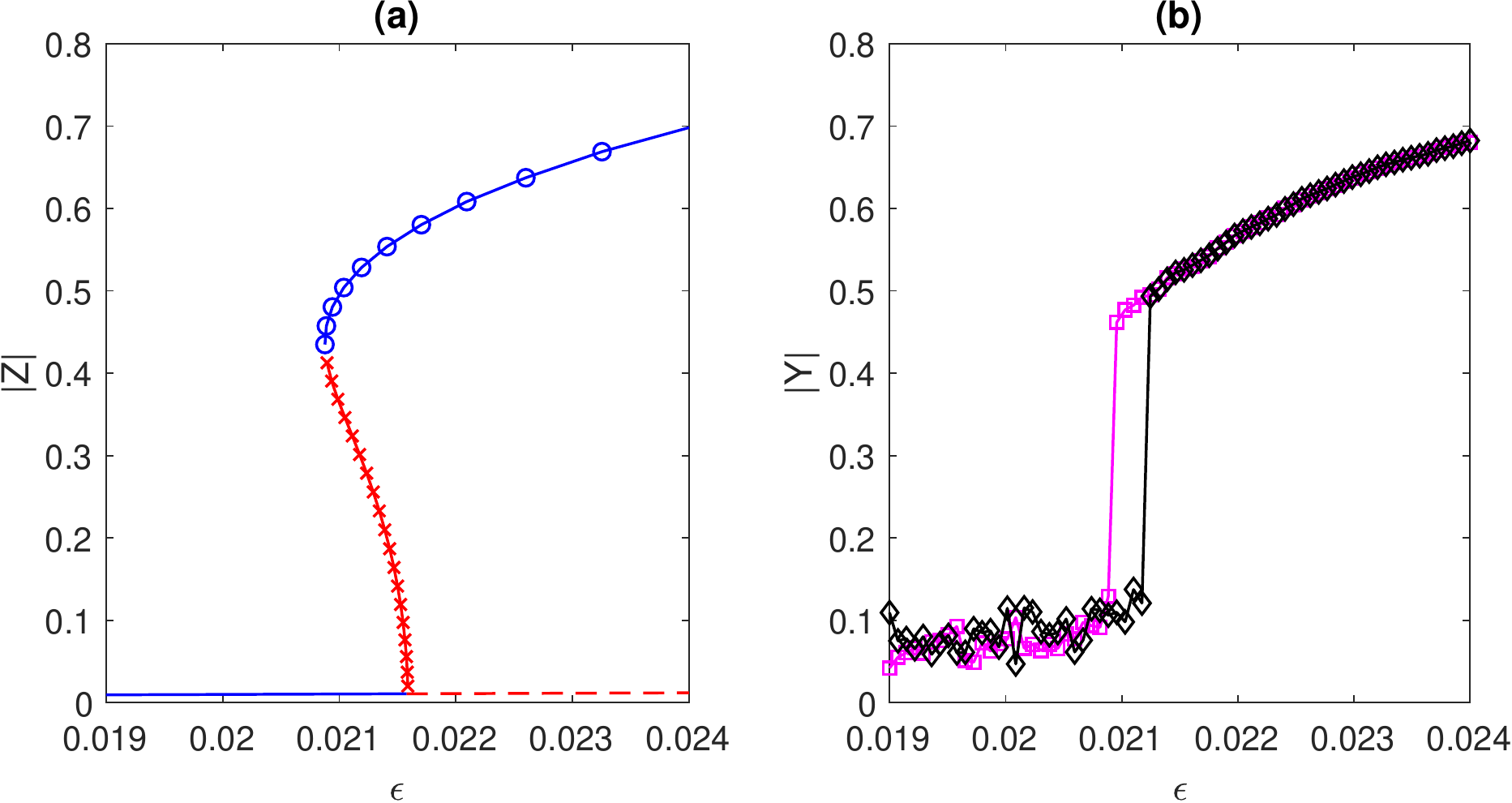}
\caption{(a): $|Z|$ for fixed point (lines) and periodic solutions (symbols) of~\eqref{eq:dbdt}.
Blue solutions are stable while red are unstable. (b): $|Y|$ for the network~\eqref{eq:dthdt}.
Black corresponds to increasing $\epsilon$ and magenta to decreasing.
Parameters: $m=100,M=400, N=2000, \sigma=0.01,\Delta=0.0005$.}
\label{fig:gau}
\end{center}
\end{figure}

We first
vary $\epsilon$ with $\sigma=0.01$. 
Results are shown in Fig.~\ref{fig:gau}, where panel (a) shows results 
from~\eqref{eq:dbdt}. For small $\epsilon$~\eqref{eq:dbdt} has a stable fixed point at which
the network is incoherent, with $|Z|$ being small.
As $\epsilon$ is increased the fixed point undergoes a subcritical Hopf bifurcation, becoming
unstable. (In the all-to-all coupled network, this Hopf bifurcation is supercritical.)
The unstable periodic orbit created in this bifurcation
is shown with red crosses and it becomes stable in a
saddle-node bifurcation. Thus there is a small range of $\epsilon$ values for which the network
is bistable. (For periodic orbits,
the quantity plotted on the vertical axis is the average over one period of $|Z(t)|$.)

Panel (b) of Fig.~\ref{fig:gau} shows 
$|Y|$ for the network~\eqref{eq:dthdt} with $\epsilon$ quasistatically
increased or decreased, using the final state of the network at one value of $\epsilon$
as the initial condition for the next value. The value plotted is the mean over 5000 time units
of $|Y(t)|$. The bistability and hysteresis is clear.
Networks of the form used in~\eqref{eq:dthdt} were created using the configuration 
model~\cite{newman03}. Both self-connections and multiple connections between oscillators
removed in a random way~\cite{laing2021}. Values of $\omega_j$ were then assigned as above, using 
Lorentzian distributions centred at target frequencies.


\begin{figure}
\begin{center}
\includegraphics[width=14cm]{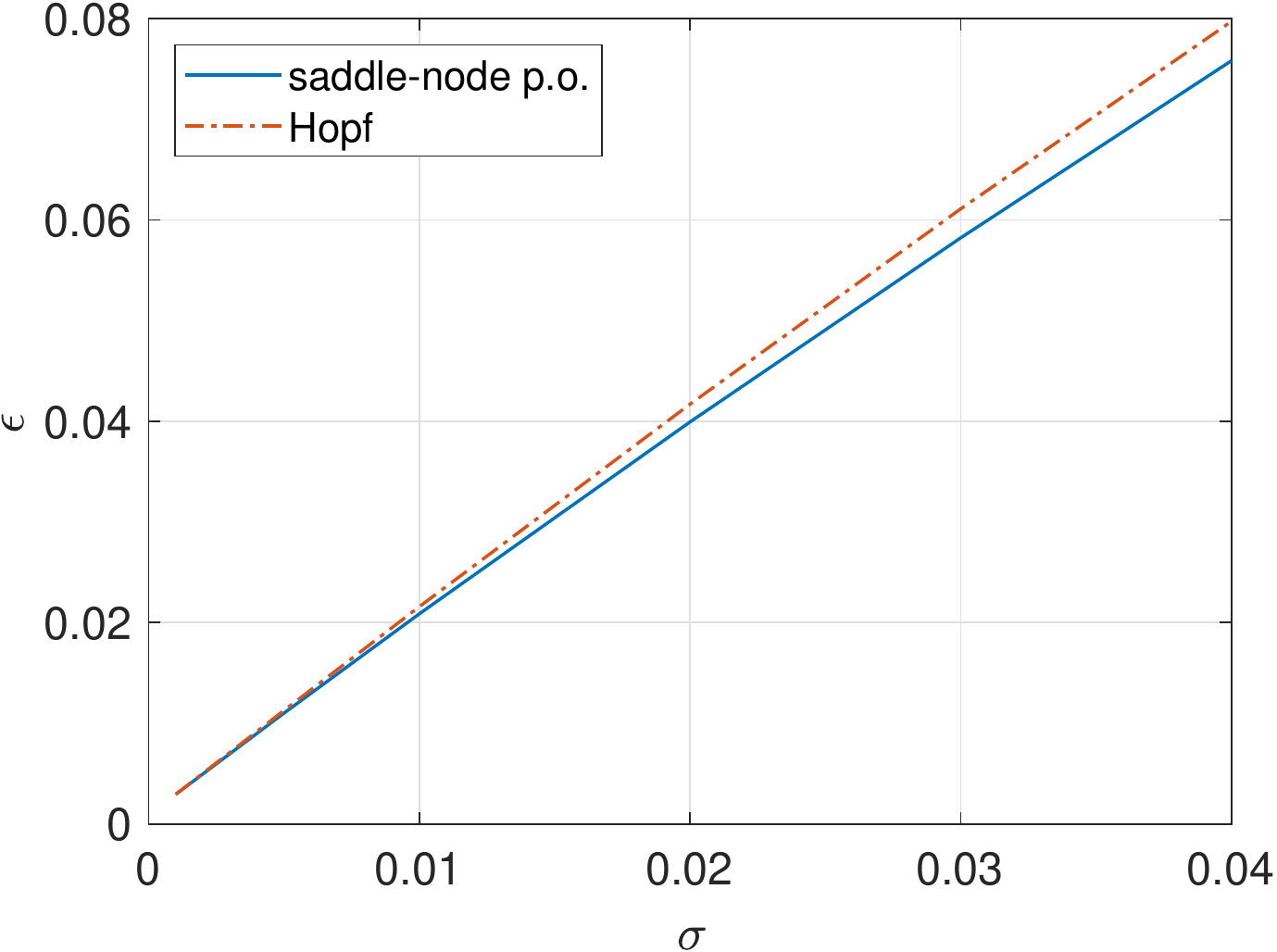}
\caption{Curves of Hopf and saddle-node bifurcations as seen in 
Fig.~\ref{fig:gau}(a) as both $\sigma$ and $\epsilon$ are varied. The network is bistable
between the curves.
Parameters: $m=100,M=400,\Delta=0.0005$.}
\label{fig:gau2p}
\end{center}
\end{figure}

Following the Hopf bifurcation 
and saddle-node bifurcation of periodic orbits shown in Fig.~\ref{fig:gau}(a)
as both $\sigma$ and $\epsilon$ are varied we obtain Fig.~\ref{fig:gau2p}. Although
the range of values
of $\epsilon$ for which the system is bistable is small, it increases as $\sigma$ is increased.
We have shown that even with very different distributions of in-degrees and intrinsic frequencies,
positively correlating them can induce ES in a directed network of Winfree oscillators. 

\begin{figure}
\includegraphics[width=.95\linewidth]{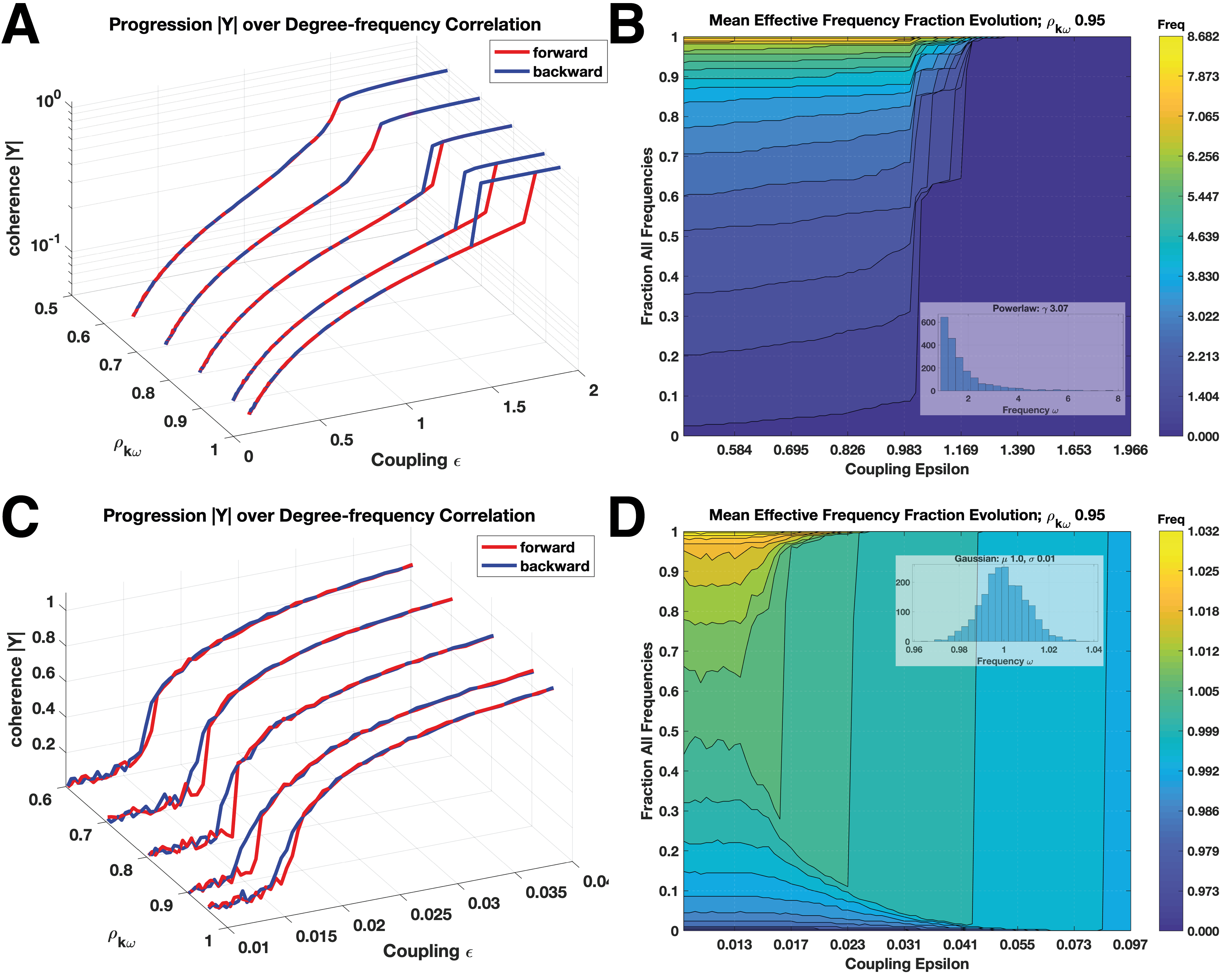}
\caption{Panels A and C: $|Y|$ as $\epsilon$ is quasistatically varied for networks with
different values of $\rho_{k\omega}$. Panels B and D: Progression of mean-effective frequency fraction distribution over increasing coupling strength $\epsilon$, for $\rho_{k\omega}$ = 0.95;
note distinct colour scales for effective frequency evolution due to underlying $\omega$ distributions (histograms, insets Panels B \& D), and distinct $\epsilon$ scales reflecting hysteresis over different coupling strengths.	
Panels A and B show results when the intrinsic frequencies are power law distributed with
an exponent of $-3.07$ and in- and out-degrees drawn from power law distribution with same exponent and neutral correlation between the in- and out-degrees. Panels C and D show results for Gaussian distributed frequencies with mean $1.0$ and $\sigma=0.01$, while in- and out-degrees are
power law distributed with exponent $-2.66$ and the in-degrees are highly correlated with 
the out-degrees ($\rho=0.89$). }
\label{fig:result_rho_progress1}
\end{figure}

The results of our numerical investigation into the effects of varying the correlation
between the $k_{in,j}$ and $\omega_j$ ($\rho_{k\omega}$) are shown in 
Fig~\ref{fig:result_rho_progress1} (panels C and D). We varied $\rho_{k\omega}$ between 0.6 and
0.95 using the Monte-Carlo degree-swapping method explained above and for each network
we quasistatically increased or decreased the coupling strength $\epsilon$ and measured
the time-averaged value of $|Y|$. This is shown in Fig~\ref{fig:result_rho_progress1}C where
we see results similar to those in Fig.~\ref{fig:gau} --- a small region of bistability.
Panel D of Fig~\ref{fig:result_rho_progress1} shows the fraction of effective frequencies 
as the coupling strength is progressively increased, at $\rho_{k\omega}=0.95$. 

 The distribution of frequencies and their fractional evolution corresponds to the mean and 
standard deviation of the Gaussian distribution, and the highest concentration of frequencies around the mean emerge dominant. However, the effects of having a large value of $\rho_{k\omega}$ for  Gaussian distributed intrinsic frequencies is minimal compared with that for power law distributed 
frequencies, considered next.

\section{Power law frequency distribution}
\label{sec:pow}

We keep the power law distribution of degrees~\eqref{eq:pk} and
 now consider the case where the target distribution frequency distribution is also power law
distributed and limited between $c$ and $C$, but with the exponent as a parameter,
i.e.
\be
   p_{\omega_0}(\omega_0)=\begin{cases} a_\omega/\omega_0^{\gamma+1} & c\leq \omega_0\leq C \\ 0 & \mbox{ otherwise} \end{cases} \label{eq:pw}
\ee 
where $a_\omega=\gamma c^\gamma C^\gamma/(C^\gamma-c^\gamma)$. As above, having created
a network we randomly choose target frequencies from the distribution~\eqref{eq:pw},
then assign the smallest target frequency to the
oscillator with the smallest in-degree, all the way up to the largest target frequency being
associated with oscillator with the largest in-degree. The actual $\omega_j$ are then chosen from a
Lorentzian with HWHM equal to $\Delta$ 
centred at the target frequency, as above. The dependence of $\omega_0$ on $k$ is then
\be
   \omega_0(k)=\widehat{p}_{\omega_0}^{-1}(\widehat{p}(k)) \label{eq:pom}
\ee
where $\widehat{p}_{\omega_0}$ is the cumulative distribution function of $p_{\omega_0}$, i.e
\be
   \widehat{p}_{\omega_0}(\omega_0)=\frac{a_\omega}{\gamma}\left(\frac{1}{c^\gamma}-\frac{1}{\omega_0^\gamma}\right) 
\ee

\subsection{Highly correlated degree and frequency}

\subsubsection{Independent degrees}
\begin{figure}
\begin{center}
\includegraphics[width=14cm]{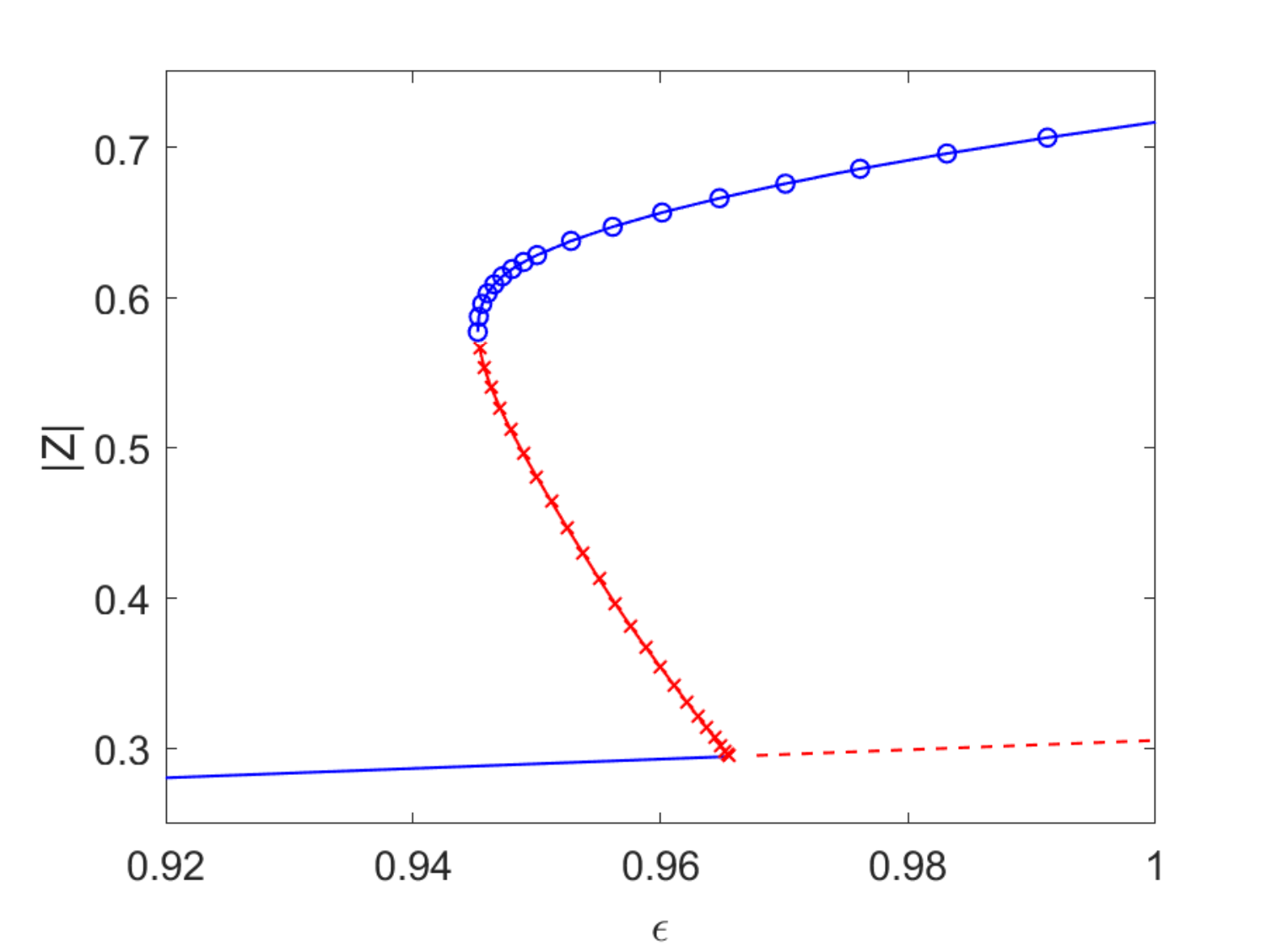}
\caption{$|Z|$ for fixed point (lines) and periodic (symbols) solutions of~\eqref{eq:dbdt}.
Blue solutions are stable while red are unstable. 
Parameters: $c=1,C=6,m=100,M=400,\gamma=2,\Delta=0.01$.}
\label{fig:powA}
\end{center}
\end{figure}

We first consider the case of independent degrees, as in Sec.~\ref{sec:gauss}. Thus
we numerically investigate~\eqref{eq:dbdt} with~\eqref{eq:Rind}, but using~\eqref{eq:pom}.
We choose $\Delta=0.01$, set $c=1,C=6$, and initially choose $\gamma=2$. 
As in Sec.~\ref{sec:gauss}, for small
$\epsilon$ the system has a stable fixed point and this becomes unstable through a subcritical
Hopf bifurcation as $\epsilon$ is increased. The results are shown in Fig.~\ref{fig:powA} where
we see the periodic orbit created in the Hopf bifurcation, giving the
same scenario as in Fig.~\ref{fig:gau}~(a).
Quasistatically increasing $\epsilon$ the solution of~\eqref{eq:dbdt} 
would jump from a fixed point to a periodic
state with amplitude significantly larger than zero. Decreasing $\epsilon$, the solution
would jump from a finite-amplitude periodic orbit to a fixed point.
(As above, for periodic orbits,
the quantity plotted on the vertical axis is the average over one period of $|Z(t)|$.)

\begin{figure}
\begin{center}
\includegraphics[width=14cm]{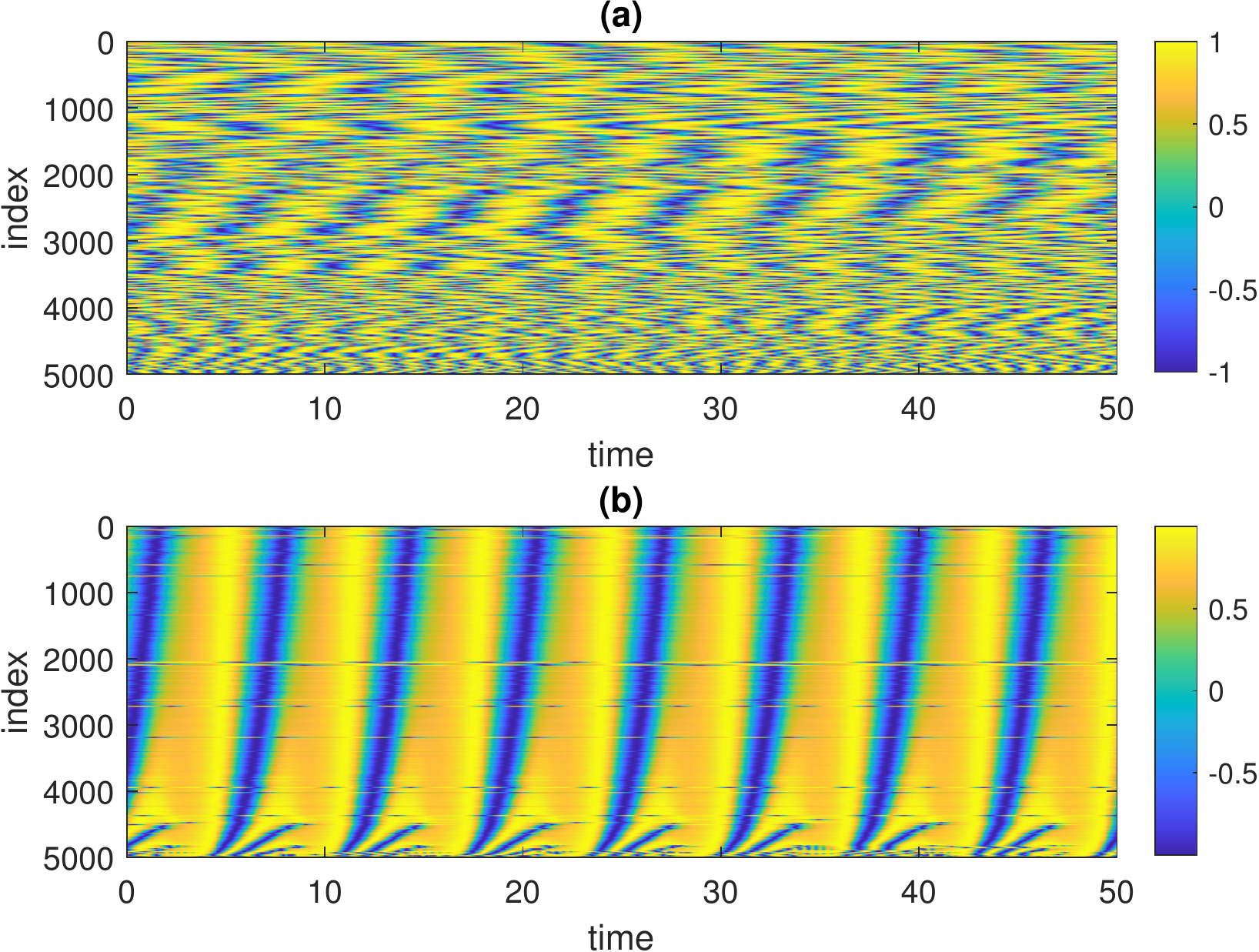}
\caption{$\sin{\theta}$ shown in colour for a simulation of~\eqref{eq:dthdt}.
The oscillators are sorted by their in-degree.
(a): $\epsilon=0.92$. (b): $\epsilon=1$.
Other parameters: $N=5000,c=1,C=6,m=100,M=400,\gamma=2,\Delta=0.01$.}
\label{fig:dyn}
\end{center}
\end{figure}

The behaviour of a particular realisation of the discrete network~\eqref{eq:dthdt} is 
slightly different, since a fixed point of~\eqref{eq:dbdt} corresponds to an incoherent
solution of~\eqref{eq:dthdt} for which $|Y|$ is not constant, having small fluctuations about
an average value. An example of such dynamics is shown in Fig.~\ref{fig:dyn}(a), with
$\epsilon=0.92$. (Other parameters have the same values as in Fig.~\ref{fig:powA}.)
 We see that most oscillators are oscillating, but with independent phases.
Similarly, a periodic solution of~\eqref{eq:dbdt} corresponds to a solution
of~\eqref{eq:dthdt} for which $|Y|$ is nearly periodic, with the vast majority of the
oscillators having the same average frequency, as shown in Fig.~\ref{fig:dyn}(b) ($\epsilon=1$). 
Some with high in-degree fire at multiples of this frequency and some are unlocked, giving a state
referred to by~\cite{ariaratnam2001} as ``partially locked hybrid states.'' See also~\cite{pazmon14}
for examples of these dynamics.

\begin{figure}
\begin{center}
\includegraphics[width=14cm]{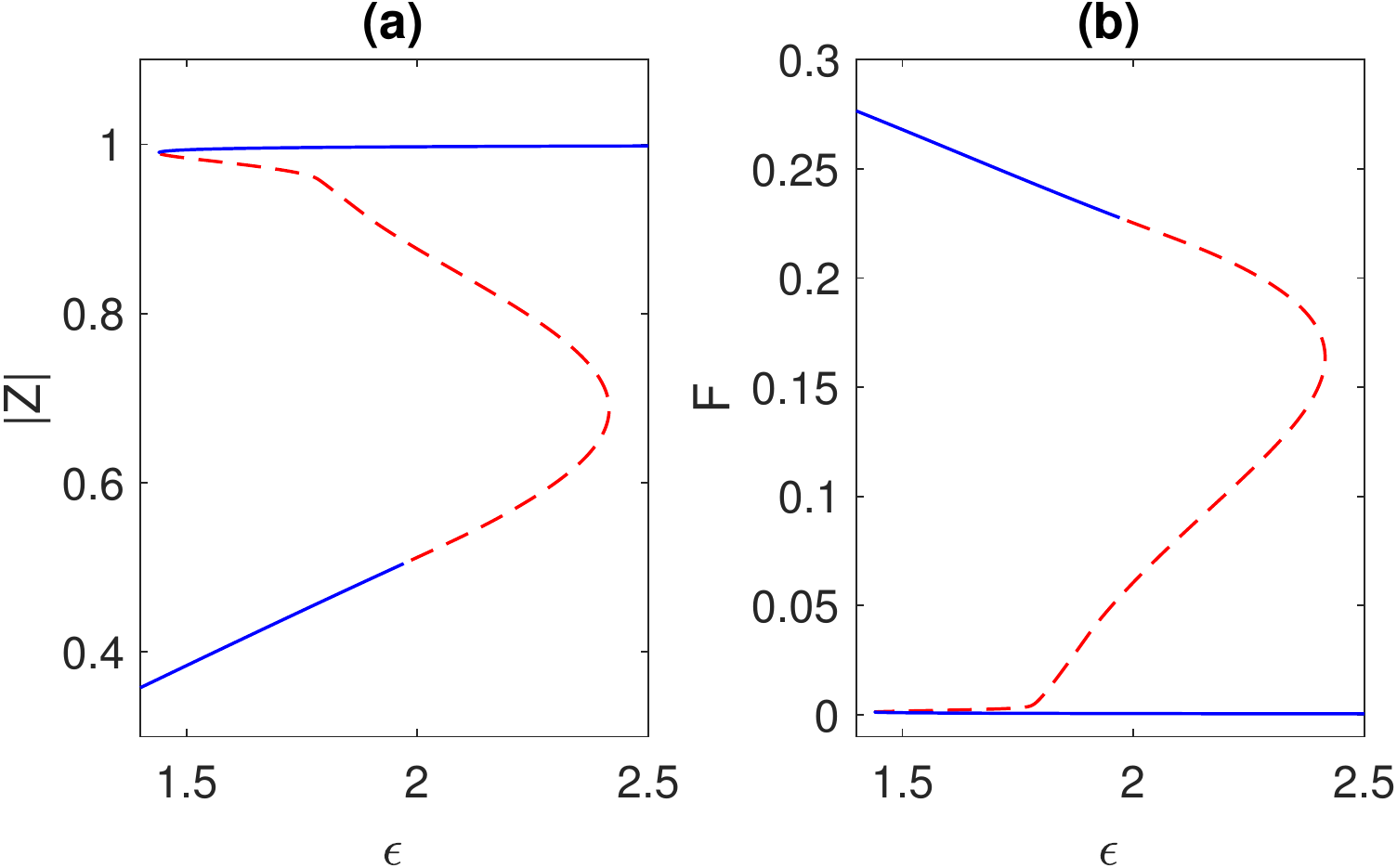}
\caption{(a) $|Z|$ and (b) mean firing rate $F$ for fixed points of~\eqref{eq:dbdt}.
Solid: stable; dashed: unstable. The bifurcation on the lower branch in (a) is a subcritical Hopf.
Parameters: $c=1,C=6,m=100,M=400,\gamma=1,\Delta=0.01$.}
\label{fig:powB}
\end{center}
\end{figure}

However for $\gamma=1$ the results are very different. 
As seen in Fig.~\ref{fig:powB}(a) the branch with the lower value of $|Z|$
still undergoes a Hopf bifurcation. Numerical integration of~\eqref{eq:dbdt}
near the bifurcation does not show a stable periodic orbit so this bifurcation
seems subcritical, creating
an unstable periodic orbit as $\epsilon$ is decreased. This orbit never becomes stable, and we hypothesise 
that it is destroyed in
a homoclinic bifurcation with the ``middle'' unstable branch.
Thus the system has no stable periodic orbits, but rather a region of
bistability between two fixed points with
different values of $|Z|$. Either increasing or decreasing
$\epsilon$ the network jumps from one fixed point to another.
Fig.~\ref{fig:powB}(b) shows the mean firing rate $F$ across the network and we see that
the branch with large $|Z|$ has small $F$ and vice versa. So even if $|Z|$ is large,
normally indicating synchronous oscillations, here it corresponds to a state 
in which most of the oscillators
are locked at an approximate fixed point, not firing. 

\begin{figure}
\begin{center}
\includegraphics[width=12cm]{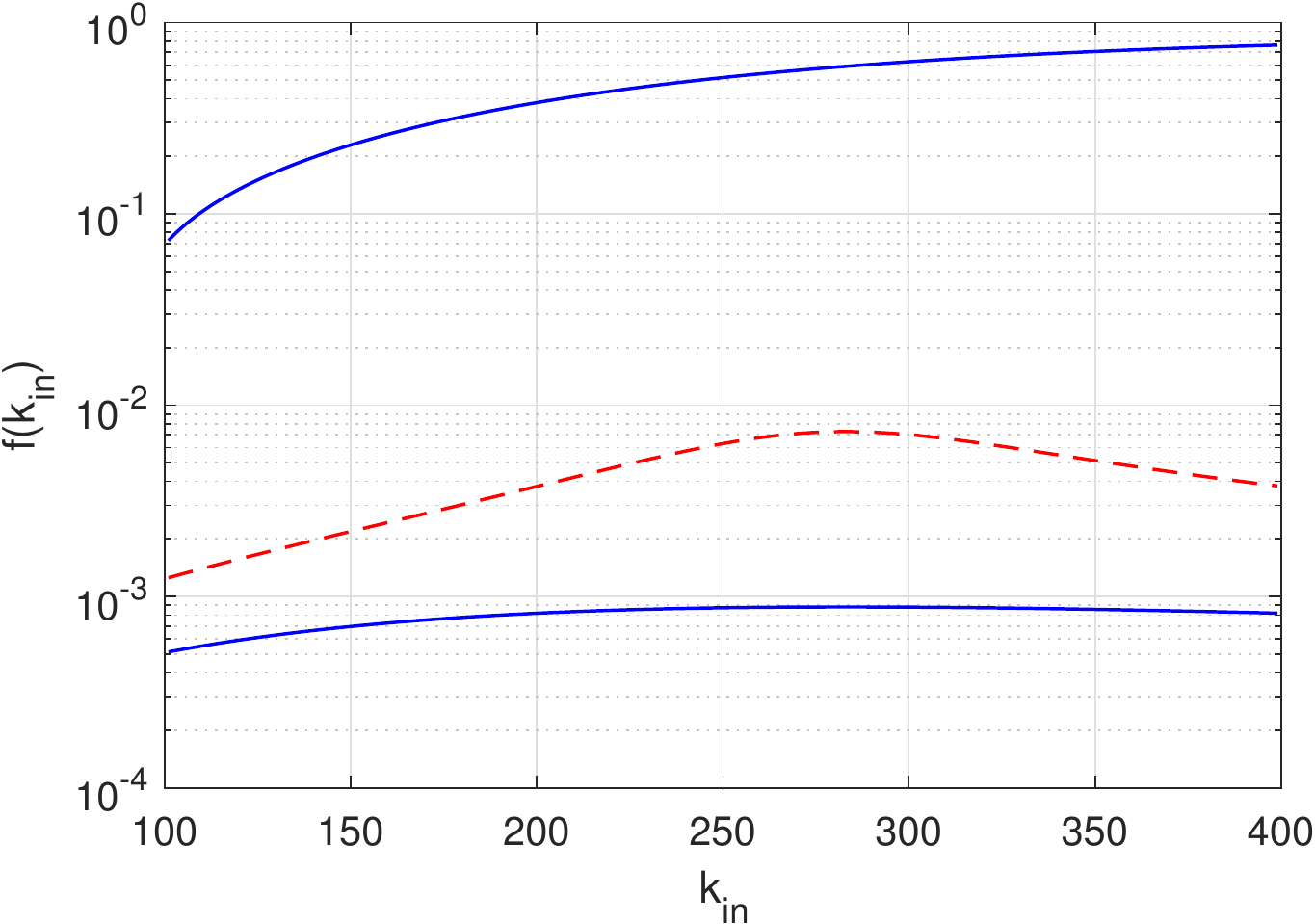}
\caption{Expected frequency for oscillators with in-degree $k_{in}$, given by~\eqref{eq:fkin},
for three coexisting steady states at $\epsilon=1.7$. See Fig.~\ref{fig:powB}.
Blue: stable solutions; red: unstable solution.
Other parameters: $c=1,C=6,m=100,M=400,\gamma=1,\Delta=0.01$.}
\label{fig:prof6}
\end{center}
\end{figure}

Fig.~\ref{fig:prof6} shows (on a logarithmic scale)
the expected frequency for oscillators with in-degree $k_{in}$, 
given by~\eqref{eq:fkin},
for three coexisting steady states in Fig.~\ref{fig:powB} at $\epsilon=1.7$. The stable solution
with highest frequencies has the lowest value of $|Z|$ and vice versa. Interestingly, for the
upper curve the frequency increases with in-degree $k_{in}$, but for the lower two curves
the maximum frequency does not occur at either extreme of the $k_{in}$ values. 

\begin{figure}
\begin{center}
\includegraphics[width=14cm]{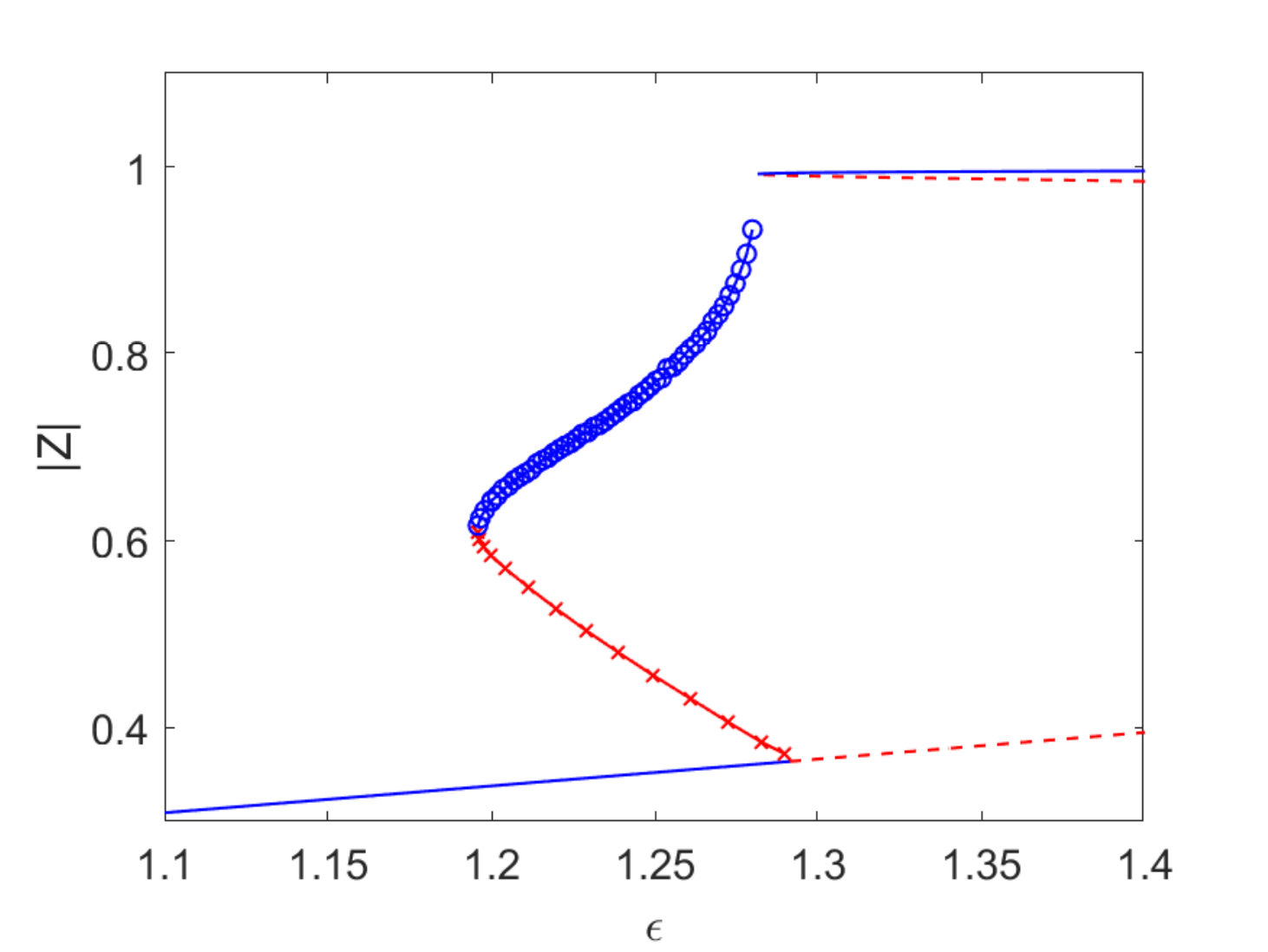}
\caption{$|Z|$ for fixed point (lines) and periodic (symbols) solutions of~\eqref{eq:dbdt}.
Blue solutions are stable while red are unstable.  The stable branch of periodic orbits
terminates at the SNIC bifurcation. 
Parameters: $c=1,C=6,m=100,M=400,\gamma=1.5,\Delta=0.01$.}
\label{fig:powC}
\end{center}
\end{figure}

A third scenario occurs for $\gamma=1.5$, as seen in Fig.~\ref{fig:powC}. The fixed point that
is stable for small $\epsilon$ becomes unstable through a subcritical Hopf bifurcation as
$\epsilon$ is increased as in Fig.~\ref{fig:powA}, but the stable periodic orbit is
destroyed in a SNIC bifurcation which occurs at a slightly lower value of $\epsilon$ than that
at which the Hopf bifurcation occurs. So if $\epsilon$ is slowly increased the network
will jump from one fixed point to another fixed point. But if $\epsilon$
is then decreased the network will switch from a fixed point to
a stable periodic orbit. This stable orbit is then destroyed in a saddle-node
bifurcation as $\epsilon$ is further decreased and the network will jump to the original
fixed point.

\subsubsection{Identical degrees}
\begin{figure}
\begin{center}
\includegraphics[width=14cm]{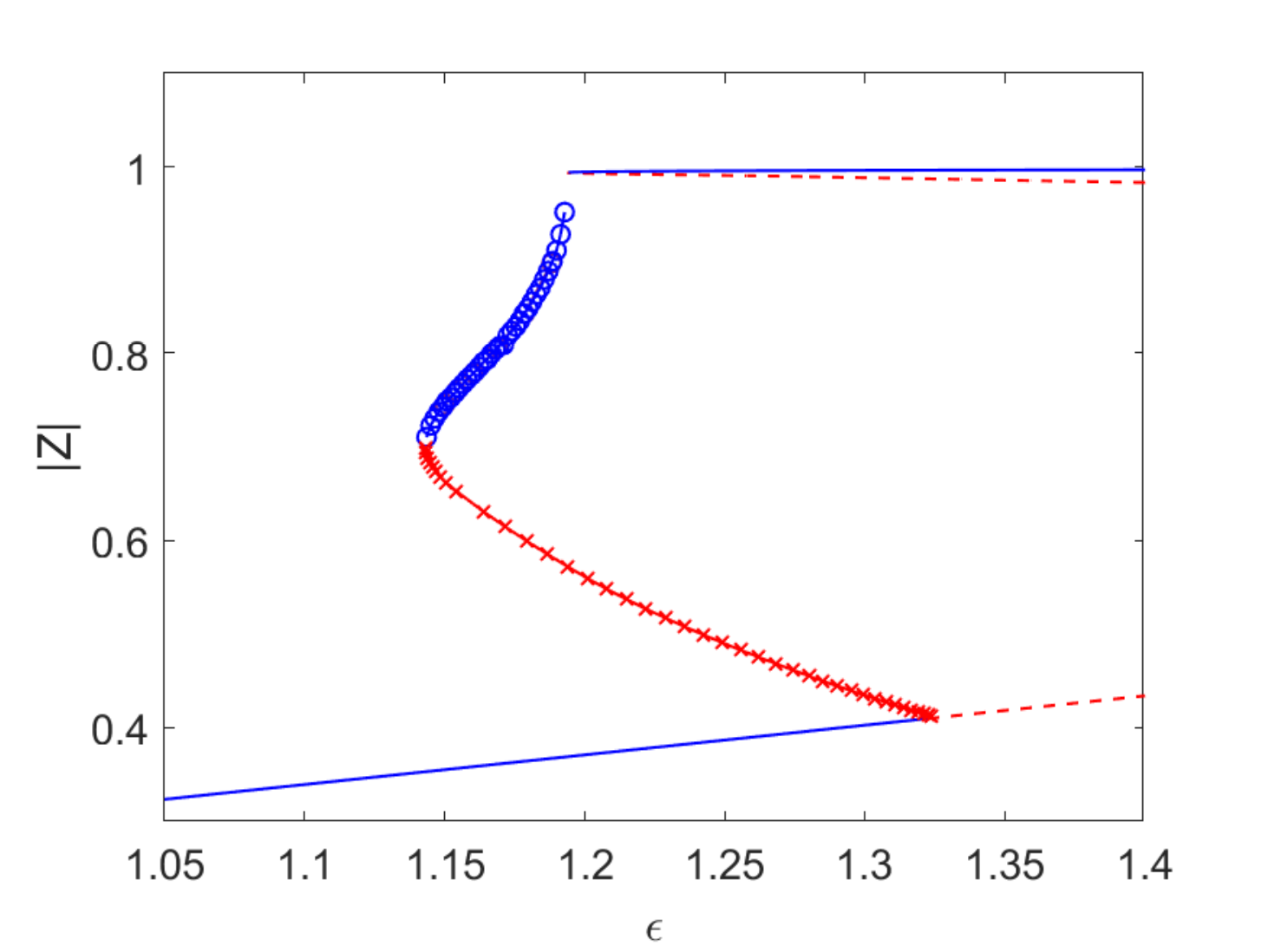}
\caption{Identical degrees.
$|Z|$ for fixed point (lines) and periodic (symbols) solutions of~\eqref{eq:dbdt}.
Blue solutions are stable while red are unstable. The stable branch of periodic orbits
terminates at the SNIC bifurcation. 
Parameters: $c=1,C=6,m=100,M=400,\gamma=2,\Delta=0.01$.}
\label{fig:idA}
\end{center}
\end{figure}

We now consider the case of identical in- and out-degrees.
We numerically investigate~\eqref{eq:dbdt} with~\eqref{eq:Rsame}, 
using the power law distribution of degrees~\eqref{eq:pk} and the
power law distribution of frequencies~\eqref{eq:pw}. The results for $\gamma=2$
are shown
in Fig.~\ref{fig:idA}. 
The scenario is qualitatively the same as in Fig.~\ref{fig:powC}, with bistability
between either two fixed points or between a fixed point and a periodic solution.

\begin{figure}
\begin{center}
\includegraphics[width=14cm]{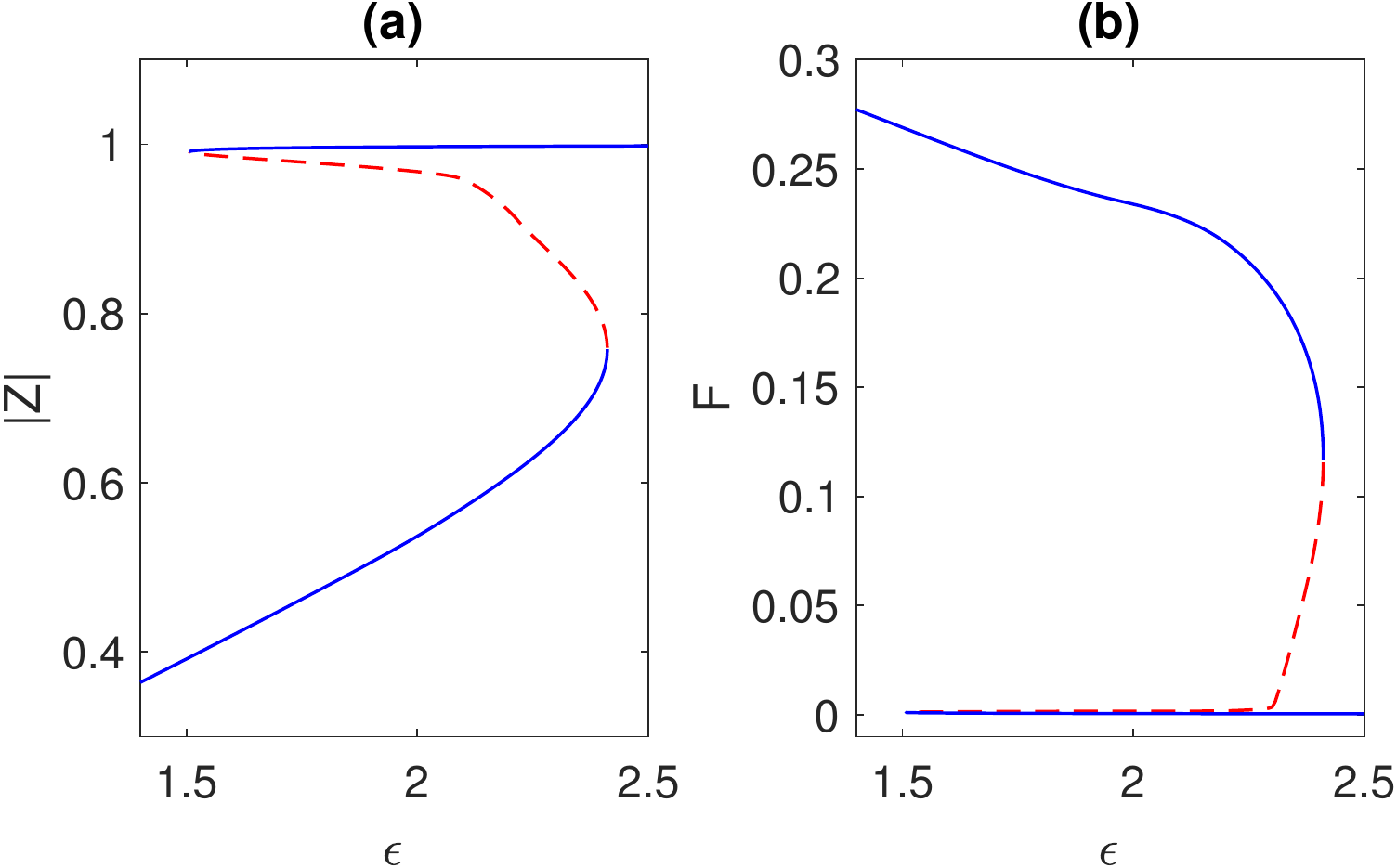}
\caption{Identical degrees.
(a) $|Z|$ and (b) mean firing rate $F$ for fixed points of~\eqref{eq:dbdt}.
Solid: stable; dashed: unstable. 
Parameters: $c=1,C=6,m=100,M=400,\gamma=1,\Delta=0.01$.}
\label{fig:idB}
\end{center}
\end{figure}

For $\gamma=1$ we obtain Fig.~\ref{fig:idB}, showing yet another scenario. Here there are
no Hopf bifurcations, only two saddle-node bifurcations of fixed points, with region
of bistability between them.

All of the results reported in this section
have been verified using simulations of the discrete network
(results not shown).

\subsubsection{Varying degree-frequency correlation}

We repeated our numerical investigation into the effects of varying degree-frequency correlations
--- which we denoted $\rho_{k\omega}$ --- generated via the Monte-Carlo swapping and permutation assembly method with power law distributed frequencies. The results are shown in Fig. \ref{fig:result_rho_progress1} panels A and B. Increasing $\rho_{k\omega}$ corresponds with the appearance and subsequent widening of the region of hysteresis, for the power law distributed $\omega$. The Gaussian distribution (Panels C and D) alternatively demonstrates hysteresis but with narrowing at the highest level of correlation and is more sensitive to the network realisation, or final assembly of the connectivity matrix $A$ given the same degree and $\omega$ sequence. Some realisations demonstrated hysteresis but others did not with all other elements identical, and compared to the power law distributed frequencies, we observed less consistency in emergent hysteresis across the multiple network realisations.  Further, the Gaussian $\omega$ results are clearly and simply noisier, partly due to the scale of plots showing hysteresis at far lower $\epsilon$ values. This is also due to the modes of synchronisation for these two $\omega$ distributions: the Gaussian with dynamic synchrony and the power law with destructive or a quiescent synchrony as observed with locking the system at fixed points of zero frequency. Note the progression of mean effective frequency distributions with increasing coupling strength (at the highest $\rho_{k\omega} = 0.95$), where the power law $\omega$ system collapses into quiescence (Fig. \ref{fig:result_rho_progress1} panel B) as opposed to the scenario for Gaussian distributed $\omega$ (Fig. \ref{fig:result_rho_progress1} panel D). Coherence frequencies emerge at levels dictated by the concentration of $\omega$ values for each distribution: power law close to zero and Gaussian close to the mean ($\mu=1.0$). 	


\section{Discussion}
\label{sec:disc}
We have considered directed networks of Winfree oscillators with truncated inverse power law
distributions of both in- and out-degrees. We considered the case of oscillators
having independent in- and out-degrees, and also the case where they are equal.
For independent degrees we examined both Gaussian and power law distributed intrinsic
frequencies, having these frequencies highly correlated with oscillators' in-degrees as a
result of sorting both sets and associating quantities of equal rank. We investigated
the effects of varying this degree of correlation. For identical
in- and out-degrees we also considered power law distributed frequencies with a positive
correlation between an oscillator's in-degree and its intrinsic frequency. We varied
both the width of the Gaussian frequency distribution (when used) and the exponent in the
power law frequency distribution and examined the transitions that occurred as the strength
of coupling within the network was varied.

In all cases shown there was an ``explosive'' transition as the coupling strength was varied
from one fixed point to another, or between a fixed point and a periodic solution. An exception occurs with lower degree-frequency correlations for the power law frequency distributed case. A variety
of scenarios were seen. This is
in contrast to the similar and much more widely studied Kuramoto model, for which
transitions are only between an incoherent fixed point and a partially synchronous periodic
solution~\cite{dsouza2019,gomez2011}. 
The range of scenarios is a result of the form of a Winfree oscillator: for strong
enough coupling an oscillator will approximately lock to a fixed point, so that even though
the measure of synchrony $|Z|$ may increase as coupling strength is increased,
this does not necessarily correspond to synchronous firing, as can be seen in Figs.~\ref{fig:powB}
and~\ref{fig:idB}.

Acknowledgements: This work was partially supported by the Marsden Fund Council from Government funding, managed by Royal Society Te Aparangi, grant number 17-MAU-054.



\end{document}